\documentclass{entcs}
\usepackage{prentcsmacro}

\usepackage{aeguill}
\usepackage{graphicx}
\usepackage{amsmath}
\usepackage{proof}
\ifpdf
  \usepackage{hyperref}
\else
  \usepackage{hyperref}
\fi

\newcommand{\C}{\textsf{C}}
\newcommand{\java}{\textsf{Java}}
\newcommand{\asdl}{\textrm{ASDL}}

\newcommand{\gom}{\textrm{GOM}}
\newcommand{\apigen}{ApiGen}
\newcommand{\aterm}{ATerm}
\newcommand{\vas}{Vas}
\newcommand{\tom}{\textrm{Tom}}
\newcommand{\SDF}{SDF}
\newcommand{\asfsdf}{\textsf{ASF+SDF}}
\newcommand{\ocaml}{OCaml}
\newcommand{\ML}{\textsf{ML}}
\newcommand{\xml}{\textsf{XML}}

\newcommand{\ie}{{i.e.}}

{\begin{tabular}{lcl}}%
{\end{tabular}}
\newcommand{\lex}[1]{{\bf {#1}}}
\newcommand{\lexstar}{\lex{\ensuremath{\boldsymbol{\ast}}}}
\newcommand{\nt}[1]{\ensuremath{\langle{\textsl{\ensuremath{#1}\/}\rangle}}}
\newcommand{\alt}{$\mid$}
\newcommand{\defg}{::=}

\newcommand{\BV}{{\ensuremath{\mathsf{BV}}}}

\newcommand{\un}{{\ensuremath{\mathord{\circ}}}}
\newcommand{\unr}{\un{\downarrow}}
\newcommand{\seqs}[1]{{\ensuremath{{\langle}#1{\rangle}}}}
\newcommand{\pars}[1]{{\ensuremath{{\lbrack}#1{\rbrack}}}}
\newcommand{\aprs}[1]{{\ensuremath{{(}#1{)}}}}
\newcommand{\cons}[1]{\{#1\}}
\newcommand{\grammareq}{\mathrel{\raise.4pt\hbox{::}{=}}}%
\newcommand{\ruleaidown}{{ai{\downarrow}}}
\newcommand{\seqr}{{q{\downarrow}}}
\newcommand{\swir}{{s}}

\def\lastname{Reilles}

\begin{document}
\begin{frontmatter}
\title{Canonical Abstract Syntax Trees}
\author{Antoine Reilles}
\address{CNRS \& LORIA\\
  Campus Scientifique, BP 239, \\
    54506 Vand{\oe}uvre-l\`es-Nancy Cedex France}
\begin{abstract}
This paper presents {\gom}, a language for describing abstract syntax trees and
generating a {\java} implementation for those trees. {\gom} includes features
allowing the user to specify and modify the interface of the data structure.
These features provide in particular the capability to maintain the internal
representation of data in canonical form with respect to a rewrite system. This
explicitly guarantees that the client program only manipulates normal forms for
this rewrite system, a feature which is only implicitly used in many
implementations.
\end{abstract}
\end{frontmatter}
\sloppy
%
\section{Introduction}
\label{sec:intro}

Rewriting and pattern-matching are of general use for describing computations
and deduction. Programming with rewrite rules and strategies has been proven
most useful for describing computational logics, transition systems or
transformation engines, and the notions of rewriting and pattern matching are
central notions in many systems, like expert systems (JRule), programming
languages based on rewriting (ELAN, Maude, OBJ) or functional programming 
 (\ML, Haskell).

In this context, we are developing the {\tom} system~\cite{MoreauRV-2003},
which consists of a language extension adding syntactic and associative
pattern matching and strategic rewriting capabilities to existing languages
like {\java}, {\C} and {\ocaml}.
This hybrid approach is particularly well-suited when describing
transformations of structured entities like trees/terms and
{\xml} documents.

One of the main originalities of this system is to be data structure
independent. This means that a \textit{mapping\/} has to be defined to
connect algebraic data structures, on which pattern matching is
performed, to low-level data structures, that correspond to the
implementation.
Thus, given an algebraic data structure definition, it is needed to implement
an efficient support for this definition in the language targeted by the
{\tom} system, as {\java} or {\C} do not provide such data structures.
Tools like {\apigen}~\cite{apigen-2003} and {\vas}, which is a human readable
language for {\apigen} input where used previously for generating such an
implementation to use with {\tom}.

However, experience showed that providing an efficient term data structure
implementation is not enough. When implementing computational logics or
transition systems with rewriting and equational matching, it is convenient to
consider terms modulo a particular theory, as identity, associativity,
commutativity, idempotency, or more problem specific
equations~\cite{marche96jsc}.

Then, it becomes crucial to provide the user of the data structure a way to
conveniently describe such rules, and to have the insurance that only chosen
equivalence class representatives will be manipulated by the program.
This need shows up in many situations. For instance when dealing
with abstract syntax trees in a compiler, and requiring constant folding
or unboxing operators protecting particular data structures.

{\gom} is a language for describing multi-sorted term algebras designed to solve
this problem. Like {\apigen}, {\vas} or {\asdl}~\cite{wang97zephyr}, its goal is to
allow the user of an imperative or object oriented language to describe
concisely the algebra of terms he wants to use in an application, and to
provide an (efficient) implementation of this algebra.

Moreover, it provides a mechanism to describe normalization functions for the
operators, and it ensures that all terms manipulated by the user of the data
structure are normal with respect to those rules.
{\gom} includes the same basic functionality as {\apigen} and {\vas}, and
ensures that the data structure implementation it provides are maximally
shared.  Also, the generated data structure implementation supports the visitor
combinator~\cite{visser-oopsla01} pattern, as the strategy language of {\tom}
relies on this pattern.

Even though {\gom} can be used in any {\java} environment, its features have been
designed to work in synergy with {\tom}. Thus, it is able to generate correct
{\tom} mappings for the data structure ({\ie} being \emph{formal
anchors\/}~\cite{KirchnerMR-PPDP2005}).
{\gom} provides a way to define computationally complex constructors for a data
structure. It also ensures those constructors are used, and that no
\emph{raw\/} term can be constructed. Private types~\cite{ocaml-manual} in the
{\ocaml} language do provide a similar functionality by hiding the type
constructors in a private module, and exporting construction functions.
However, using private types or normal types is made explicit to the user,
while it is fully transparent in {\gom}.
{MOCA}, developed by Fr\'ed\'eric Blanqui and Pierre Weis is a tool that
implements normalization functions for theories like associativity or
distributivity for {\ocaml} types. It internally uses private types to
implement those normalization functions and ensure they are used, but could
also provide such an implementation for {\gom}.

The rest of the paper is organized as follows: in Section~\ref{sec:tom}, to
motivate the introduction of {\gom}, we describe the {\tom} programming
environment and its facilities.
Section~\ref{sec:gom} presents the {\gom} language, its semantics and
some simple use cases.
After presenting how {\gom} can cooperate with {\tom} in
Section~\ref{sec:interact},
%
we expose 
in Section~\ref{sec:structure} the example of a prover for the calculus of
structures~\cite{Gug02}
%
%
showing how the combination of {\gom} and {\tom} can help
producing a reliable and extendable implementation for a complex system.
We conclude with summary and discussions in Section~\ref{sec:conclusion}.
%
\section{The {\tom} language}
\label{sec:tom}

{\tom} is a language extension which adds pattern matching primitives to
existing imperative languages. Pattern-matching is directly related to the
structure of objects and therefore is a very natural programming language
feature, commonly found in functional languages. This is particularly
well-suited when describing various transformations of structured entities
like, for example, trees/terms, hierarchized objects, and {\xml} documents.

The main originality of the {\tom} system is its language and data-structure
independence. From an implementation point of view, it is a compiler which
accepts different \emph{native languages\/} like {\C} or {\java} and whose
compilation process consists in translating the matching constructs into the
underlying native language.

It has been designed taking into account experience about efficient compilation
of rule-based systems~\cite{KirchnerM-JFP2001}, and allows the definition of
rewriting systems, rewriting rules and strategies.
For an interested reader, design and implementation issues related to {\tom}
are presented in~\cite{MoreauRV-2003}.

{\tom} is based on the notion of formal anchor presented
in~\cite{KirchnerMR-PPDP2005}, which defines a mapping between the algebraic
terms used to express pattern matching and the actual objects the underlying
language manipulates.
Thus, it is data structure independent, and customizable for any term
implementation.

For example, when using {\java} as the host language, the
sum of two integers can be described in {\tom} as follows:
{\begin{small}
\begin{verbatim}
  Term plus(Term t1, Term t2) {
    %match(Nat t1, Nat t2) {
      x,zero   -> { return x; }
      x,suc(y) -> { return suc(plus(x,y)); } 
    }
  }
\end{verbatim}
\end{small}}
Here the definition of \verb|plus| is specified functionally, but the function
\verb|plus| can be used as a {\java} function to perform addition.
\verb|Nat| is the algebraic sort {\tom} manipulates, which is mapped to {\java}
objects of type \verb|Term|. The mapping between the actual object \verb|Term|
and the algebraic view \verb|Nat| has to be provided by the user.

The language provides support for matching modulo sophisticated
theories. For example, we can specify a matching modulo associativity
and neutral element (also known as list-matching) that is particularly
useful to model the exploration of a search space and to perform list
or {\xml} based transformations.
To illustrate the expressivity of list-matching we can define the search of a
\verb|zero| in a list as follows:

{\begin{small}
\begin{verbatim}
  boolean hasZero(TermList l) {
    %match(NatList l) {
      conc(X1*,zero,X2*) -> { return true; }
    }
    return false;
  }
\end{verbatim}
\end{small}}
In this example, \emph{list variables}, annotated by a \verb|*| should be
instantiated by a (possibly empty) list. Given a list, if a solution to the
matching problem exists, a \verb|zero| can be found in the list and the
function returns \verb|true|, \verb|false| otherwise, since no \verb|zero| can
be found.

Although this mechanism is simple and powerful, it requires a lot of work to
implement an efficient data structure for a given algebraic signature, as well
as to provide a \emph{formal anchor} for the abstract data structure. Thus
we need a tool to generate such an efficient implementation from a given
signature. This is what tools like \apigen~\cite{apigen-2003} do.

However, {\apigen} itself only provides a tree implementation, but does not
allow to add behavior and properties to the tree data structure, like defining
ordered lists, neutral element or constant propagation in the context of a
compiler manipulating abstract syntax tree. Hence the idea to define a new
language that would overcome those problems.
%
\section{The {\gom} language}
\label{sec:gom}

We describe here the {\gom} language and its syntax, and present an example
data-structure description in \gom.
We first show the basic functionality of {\gom}, which is to provide an
efficient implementation in {\java} for a given algebraic signature.
We then detail what makes {\gom} suitable for efficiently implement normalized
rewriting~\cite{marche96jsc}, and how {\gom} allows us to write any
normalization function.

\subsection{Signature definitions}

An algebraic signature describes how a tree-like data structure should be
constructed. Such a description contains \emph{Sorts\/} and \emph{Operators}.
\emph{Operators\/} define the different node shapes for a certain
\emph{sort\/} by their name and the names and sorts of their children.
Formalisms to describe such data structure definitions include
{\apigen}~\cite{apigen-2003}, {\xml} Schema, {\ML} types, and
{\asdl}~\cite{wang97zephyr}.

To this basic signature definition, we add the notion of \emph{module\/} as a
set of sorts. This allows to define new signatures by composing existing
signatures, and is particularly useful when dealing with huge signatures, as
can be the abstract syntax tree definition of a compiler.
Figure~\ref{fig:lightsyntax} shows a simplified syntax for {\gom} signature
definition language. In this syntax, we see that a module can import existing
modules to reuse its sorts and operators definitions. Also, each module
declares the sorts it defines with the \lex{sorts} keyword, and declares
operators for those sorts with productions.
\begin{figure}
\begin{center}
\def\arraystretch{1.2}
\begin{tabular}{lcl}
  \nt{Gom} & \defg\ & \nt{Module}\\
  \nt{Module} & \defg\ & \lex{module} \nt{ModuleName} \\
              & & [\nt{Imports}] \nt{Grammar} \\
  \nt{Imports} & \defg\ & \lex{imports} (\nt{ModuleName})*\\
  \nt{Grammar} & \defg\ & \lex{sorts} (\nt{SortName})*\\
  &  & \lex{abstract syntax} (\nt{Production})*\\
  \nt{Production} & \defg\ & \nt{Symbol}[\lex{ (}\nt{Field}(\lex{,}\nt{Field})* \lex{)}] \lex{->} \nt{SortName}\\
  & \alt\ & \nt{Symbol} \lex{ (} \nt{SortName} {\lexstar} \lex{)} \lex{->} \nt{SortName}\\
  \nt{Field} & \defg\ & \nt{SlotName} \lex{:} \nt{SortName}\\  
  \nt{ModuleName} & \defg\ & \nt{Identifier}\\
  \nt{SortName} & \defg\ & \nt{Identifier}\\
  \nt{Symbol} & \defg\ & \nt{Identifier}\\
  \nt{SlotName} & \defg\ & \nt{Identifier}\\
\end{tabular}
\end{center}
\caption{Simplified {\gom} syntax}
\label{fig:lightsyntax}
\end{figure}
This syntax is strongly influenced by the syntax of {\SDF}~\cite{BHKO02}, but
simpler, since it intends to deal with abstract syntax trees, instead of parse
trees. One of its peculiarities lies in the productions using the {\lexstar} symbol,
defining variadic operators. The notation {\lexstar} is the same as 
in~\cite[Section~2.1.6]{JouannaudDEA2003} for a similar construction, and can be
seen as a family of operators with arities in $[0,\infty[$.

We will now consider a simple example of {\gom} signature for booleans:
{\begin{small}
\begin{verbatim}
module Boolean
  sorts Bool
  abstract syntax
    True                   -> Bool
    False                  -> Bool
    not(b:Bool)            -> Bool
    and(lhs:Bool,rhs:Bool) -> Bool
    or(lhs:Bool,rhs:Bool)  -> Bool
\end{verbatim}
\end{small}}
From this description, {\gom} generates a {\java} class hierarchy where to each
sort corresponds an abstract class, and to each operator a class extending this
\emph{sort\/} class. The generator also creates a factory class for each module
 (in this example, called \verb|BooleanFactory|), providing the user a single
entry point for creating objects corresponding to the algebraic terms.

Like {\apigen} and {\vas}, {\gom} relies on the
{\aterm}~\cite{vandenbrand00efficient} library, which provides an efficient
implementation of unsorted terms for the {\C} and {\java} languages, as a basis
for the generated classes. The generated data structure can then be
characterized by strong typing (as provided by the \emph{Composite\/} pattern
used for generation) and maximal subterm sharing. Also, the generated class
hierarchy does provide support for the visitor combinator
pattern~\cite{visser-oopsla01}, allowing the user to easily define arbitrary
tree traversals over {\gom} data structures using high level constructs
(providing congruence operators).

\subsection{Canonical representatives}
\label{sub:canon}

When using abstract data types in a program, it is useful to also define a
notion of canonical representative, or ensure some invariant of the structure.
This is particularly the case when considering an equational theory associated
to the terms of the signature, such as associativity, commutativity or neutral
element for an operator, or distributivity of one operator over another one.

Considering our previous example with boolean, we can consider the De Morgan rules
as an equational theory for booleans. De Morgan's laws state $\overline{A \vee
B} = \overline{A} \wedge \overline{B}$ and $\overline{A \wedge B} =
\overline{A} \vee \overline{B}$.
We can orient those equations to get a confluent and terminating rewrite
system, suitable to implement a normalization system, where only boolean atoms
are negated. We can also add a rule for removing duplicate negation.
We obtain the system:
{\vspace{-2ex}
\[
\def\arraystretch{1.2}
\begin{array}{lll}
\overline{A \vee B} & \rightarrow & \overline{A} \wedge \overline{B} \\
\overline{A \wedge B}  & \rightarrow & \overline{A} \vee \overline{B} \\
\overline{\overline{A}} & \rightarrow & A \\
\end{array}
\vspace{-2ex}
\]
}
{\gom}'s objective is to provide a low level system for implementing such
normalizing rewrite systems in an efficient way, while giving the user control
on how the rules are applied. 
To achieve this goal, {\gom} provides a \emph{hook\/} mechanism, allowing to
define arbitrary code to execute before, or replacing the original construction
function of an operator. This code can be any {\java} or {\tom} code, allowing
to use pattern matching to specify the normalization rules.
To allow \emph{hooks\/} definitions, we add to the {\gom} syntax the
definitions for \emph{hooks}, and add \nt{OpHook} and \nt{FactoryHook} to the
productions:
\begin{center}
\def\arraystretch{1.2}
\begin{tabular}{lcl}
  \nt{FactoryHook} & \defg\ & \lex{factory \{} \nt{TomCode} \lex{\}}\\
  \nt{OpHook} & \defg\ & \nt{Symbol} \lex{:} \nt{Operation} \lex{\{} \nt{TomCode} \lex{\}}\\
  \nt{Operation} & \defg\ & \nt{OperationType} \lex{ (} (\nt{Identifier})* \lex{)}\\
  \nt{OperationType} & \defg\ & \lex{make} \alt\ \lex{make\_before} \alt\ \lex{make\_after} \\
  \multicolumn{3}{r}{{\alt} \lex{make\_insert} \alt\ \lex{make\_after\_insert} \alt\ \lex{make\_before\_insert}}\\
  \nt{TomCode} & \defg\ & \nt{\ldots}
\end{tabular}
\end{center}
A \emph{factory hook\/} \nt{FactoryHook} is attached to the module, and allows
to define additional {\java} functions. We will see in
Section~\ref{sub:invariant} an example of use for such a \emph{hook}.
An \emph{operator hook\/} \nt{OpHook} is attached to an operator definition,
and allows to extend or redefine the construction function for this operator.
Depending on the \nt{OperationType}, the hook redefines the construction
function (\lex{make}), or insert code before (\lex{make\_before}) or after
(\lex{make\_after}) the construction function. Those \emph{hooks\/} take as
many arguments as the operator they modify has children.
We also define operation types with an appended \lex{insert}, used for
variadic operators. Those hooks only take two arguments, when the operator they
apply to is variadic, and allow to modify the operation of adding one element
to the list of arguments of a variadic operator.

Such \emph{hooks\/} can be used to define the boolean normalization system:
\begin{verbatim}
module Boolean
  sorts Bool
  abstract syntax
    True                   -> Bool
    False                  -> Bool
    not(b:Bool)            -> Bool
    and(lhs:Bool,rhs:Bool) -> Bool
    or(lhs:Bool,rhs:Bool)  -> Bool
    not:make(arg) {
      %match(Bool arg) {
        not(x)   -> { return `x; }
        and(l,r) -> { return `or(not(l),not(r)); }
        or(l,r)  -> { return `and(not(l),not(r)); }
      }
      return `make_not(arg);
    }
\end{verbatim}
We see in this example that it is possible to use {\tom} in the \emph{hook\/}
definition, and to use the algebraic signature being defined in {\gom} in the
\emph{hook\/} code. This lets the user define \emph{hooks\/} as rewriting
rules, to obtain the normalization system. The signature in the case of {\gom}
is extended to provide access to the default construction function of an
operator. This is done here with the \verb|make_not(arg)| call.

When using the \emph{hook\/} mechanism of {\gom}, the user has to ensure that
the normalization system the hooks define is terminating and confluent, as it
will not be enforced by the system. Also, combining hooks for different
equational theories in the same signature definition can lead to non confluent
systems, as combining rewrite systems is not a straightforward task.

However, a higher level strata
providing completion to compute normalization functions from their equational
definition, and allowing to combine theories and rules could take advantage of
{\gom}'s design to focus on high level tasks, while getting maximal subterm
sharing, strong typing of the generated code and \emph{hooks\/} for
implementing the normalization functions from the {\gom} strata.
{\gom} can then be seen as a reusable component, intended to be used as a tool
for implementing another language (as {\apigen} was used as basis for
{\asfsdf}~\cite{dejong2004}) or as component in a more complex architecture.
%
\section{The interactions between {\gom} and {\tom}}
\label{sec:interact}

\begin{figure}[t]
\begin{center}
\includegraphics[width=.65\textwidth]{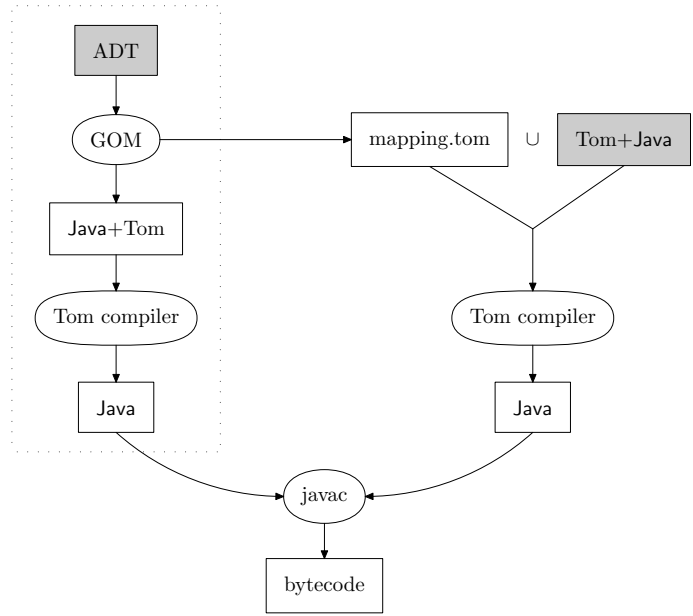}
\caption{The interaction between {\tom} and {\gom}}
\label{fig:interaction}
\end{center}
\end{figure}

The {\gom} tool is best used in conjunction with the {\tom} compiler.
{\gom} is used to provide an implementation for the abstract data type to be
used in a {\tom} program. The {\gom} data structure definition will also
contain the description of the invariants the data structure has to preserve,
by the mean of \emph{hooks}, such that it is ensured the {\tom} program will
only manipulate terms verifying those invariants.
Starting from an input datatype signature definition, {\gom} generates an
implementation in {\java} of this data structure (possibly using {\tom}
internally) and also generates an anchor for this data structure implementation
for {\tom} (See Figure~\ref{fig:interaction}).
The users can then write code using the match construct 
on the generated mapping and {\tom} compiles this to plain {\java}.
The dashed box represents the part handled by the {\gom} tool, while the grey
boxes highlight the source files the user writes.
The generated code is characterized by strong typing combined with a generic
interface and by maximal sub-term sharing for memory efficiency and fast
equality checking, as well as the insurance the hooks defined for the data
structure are always applied, leading to canonical terms. Although it is
possible to manually implement a data structure satisfying those constraints,
it is difficult, as all those features are strongly interdependent.
Nonetheless, it is then very difficult to let the data structure evolve when
the program matures while keeping those properties, and keep the task of
maintaining the resulting program manageable.

In the following example, we see how the use of {\gom} for the data structure
definition and {\tom} for expressing both the invariants in {\gom} and the
rewriting rules and strategy in the program leads to a robust and reliable
implementation for a prover in the structure calculus.
%
\section{A full example: the structure calculus}
\label{sec:structure}

We describe here a real world example of a program written using {\gom} and
{\tom} together. We implement a prover for the calculus of structure~\cite{Gug02}
where some rules are promoted to the level of data structure invariants,
allowing a simpler and more efficient implementation of the calculus rules.
Those invariants and rules have been shown correct with respect to the original
calculus, leading to an efficient prover that can be proven correct.
Details about the correctness proofs and about the proof search
strategy can be found in~\cite{KahramanogullariMR-SD05}.
We concentrate here on the implementation using {\gom}.

\subsection{The approach}
When building a prover for a particular logic, and in particular for the system
{\BV} in the structure calculus, one needs to refine the strategy of applying
the calculus rules. This is particularly true with the calculus of structure,
because of deep inference, non confluence of the calculus and
associative-commutative structures.

We describe here briefly the system {\BV}, to show how {\gom} and {\tom} can
help to provide a robust and efficient implementation of such a system.

Atoms in {\BV} are denoted by $a,b,c,\ldots$
Structures are denoted by $R, S, T,\ldots$  and  generated by
\[
  S \grammareq \un  \mid a \mid
 \seqs{\,\underbrace{S;\dots;S}_{{}>0}\,} \mid
  \pars{\,\underbrace{S,\dots,S}_{{}>0}\,} \mid
  \aprs{\,\underbrace{S,\dots,S}_{{}>0}\,} \mid \overline{S}
\]
where $\un$, the \emph{unit}, is not an atom.
$ \seqs{S;\dots;S}$ is called a \emph{seq structure},
$ \pars{S,\dots,S}$ is called a \emph{par structure}, and
$ \aprs{S,\dots,S}$ is called a \emph{copar structure}, $\overline{S}$
is the \emph{negation\/} of the structure $S$.
A structure $R$ is called a \emph{proper par structure\/} if
$R = \pars{R_1, R_2}$ where $R_1 \neq \un$ and $R_2 \neq \un$.
A \emph{structure context}, denoted as in  $S\cons{\;\;}$,
is a structure with a hole. We use this notation to express the deduction rules
for system {\BV}, and will omit context braces when there is no ambiguity.

The rules for system {\BV} are simple, provided some equivalence relations on
{\BV} terms. The seq, par and copar structures are associative, par and copar
being commutative too. Also, {\un} is a neutral element for seq, par and copar
structures, and a seq, par or copar structure with only one substructure is
equivalent to its content. Then the deduction rules for system {\BV} can be
expressed as in Figure~\ref{fig:BV}.

{
\newcommand{\rulebox}[1]{\mbox{$#1$}}
\begin{figure}[tb]
  \begin{center}
    \fbox{
      \parbox{0.9\textwidth}{
$$
\begin{array}{ll}
  \infer[\unr]{\un}{\phantom{S\pars{}}}
    \qquad
  \infer[\ruleaidown]{S\pars{a,\overline{a}}}{S\cons{\un}}
    \qquad
  \infer[\swir]{S\pars{\aprs{R,U},T}}{S\aprs{\pars{R,T},U}}
    \qquad
   \infer[\seqr]{S\pars{\seqs{R;T},\seqs{U;V} }}
        {S\seqs{\pars{R, U};\pars{T, V} }}\\
\vspace{-10mm}
\end{array}
$$
      }   }
    \vspace{-2mm}
    \caption{System {\BV}}\label{fig:BV}
  \end{center}
\end{figure}
}
Because of the contexts in the rules, the corresponding rewriting rules can be
applied not only at the top of a structure, but also on each subterm of a
structure, for implementing deep inference. Deep inference then, combined with
associativity, commutativity and {\un} as a neutral element for seq, par and
copar structures leads to a huge amount of non-determinism in the calculus.
A structure calculus prover implementation following strictly this
description will have to deal with this non-determinism, and handle a huge
search space, leading to inefficiency~\cite{Kah:ESSLLI04}.

The approach when using {\gom} and {\tom} will be to identify canonical
representatives, or preferred representatives for equivalence classes,
and implement the normalization for structures leading to the selection of the
canonical representative by using {\gom}'s \emph{hooks}.
This process requires to define the data structure first, and then define the
normalization. This normalization will make sure all units {\un} in seq, par
and copar structures are removed, as {\un} is a neutral for those structures. We
will also make sure the manipulated structures are \emph{flattened}, which
corresponds to selecting a canonical representative for the associativity of
seq, par and copar, and also that subterms of par and copar structures are
ordered, taking a total order on structures, to take commutativity into account.

When implementing the deduction rule, it will be necessary to take into account
the fact that the prover only manipulates canonical representatives. This leads
to simpler rules, and allow some new optimizations on the rules to be
performed.

\subsection{The data structure}

We first have to give a syntactic description of the structure data-type the
{\BV} prover will use, to provide an object representation for the \emph{seq},
\emph{par\/} and \emph{copar\/} structures ($\seqs{R;T}$, $\pars{R,T}$ and
$\aprs{R,T}$). In our implementation, we considered these constructors as
unary operators which take a \emph{list of structures\/} as argument.
Using {\gom}, the considered data structure can be described by the following
signature:
\begin{verbatim}
module Struct
  imports 
  public
    sorts Struc StrucPar StrucCop StrucSeq
  abstract syntax
    o -> Struc
    a -> Struc
    b -> Struc
    c -> Struc
    d -> Struc
    ...other atom constants
    neg(a:Struc) -> Struc
    concPar( Struc* )  -> StrucPar
    concCop( Struc* )  -> StrucCop
    concSeq( Struc* )  -> StrucSeq
    cop(copl:StrucCop) -> Struc
    par(parl:StrucPar) -> Struc
    seq(seql:StrucSeq) -> Struc
\end{verbatim}
To represent structures, we define first some constant
atoms. Among them, the \verb|o| constant will be used to represent the unit
{\un}. The \verb|neg| operator builds the negation of its argument.
The grammar rule \texttt{par(StrucPar) -> Struc} defines a
unary operator \verb|par| of sort \texttt{Struc} which takes a
\texttt{StrucPar} as unique argument.
Similarly, the rule \texttt{concPar(Struc*) -> StrucPar} defines the
\texttt{concPar} operator of sort \texttt{StrucPar}.
The syntax \texttt{Struc*} indicates that \texttt{concPar}
is a \emph{variadic-operator\/} which takes an indefinite number of
\texttt{Struc} as arguments.
Thus, by combining \texttt{par} and \texttt{concPar} it becomes possible to
represent the structure $\pars{a,\pars{b,c}}$ by \texttt{par(concPar(a,b,c))}.
Note that this structure is flattened, but with this description, we could also
use nested \texttt{par} structures, as in
\verb|par(concPar(a,par(concPar(b,c))))| to represent this structure.
$\aprs{R,T}$ and $\seqs{R;T}$ are represented in a similar way, using
\texttt{cop, seq}, \texttt{concCop}, and \texttt{concSeq}.

\subsection{The invariants, and how they are enforced}
\label{sub:invariant}

So far, we can manipulate objects, like \texttt{par(concPar())}, which do not
necessarily correspond to intended structures. It is also possible to have
several representations for the same structure. Hence, \texttt{par(concPar(a))}
and \texttt{cop(concCop(a))} both denote the structure~\texttt{a}, as $\seqs{R}
\approx \pars{R} \approx \aprs{R} \approx R$.

Thus, we define the canonical (prefered) representative by ensuring that
\begin{itemize}
\item $\pars{}$, $\seqs{}$ and $\aprs{}$ are reduced when containing
  only one sub-structure:\\
  $par(concPar(x)) \rightarrow x$
\item nested structures are flattened, using the rule:
\vspace{-5mm}
\[
\begin{array}{c}
par(concPar(a_1,\ldots,a_i,par(concPar(x_1,\ldots,x_n)),b_1,\ldots,b_j)) \\
\rightarrow par(concPar(a_1,\ldots,a_i,x_1,\ldots,x_n,b_1,\ldots,b_j))
\end{array}
\]
\vspace{-6mm}
\item subterms are sorted (according to a given total lexical order~$<$):\\
  $concPar(\ldots,x_i,\ldots,x_j,\ldots) \rightarrow
  concPar(\ldots,x_j,\ldots,x_i,\ldots)$ if $x_j<x_i$.
\end{itemize}
This notion of canonical form allows us to efficiently check if two
terms represent the same structure with respect to commutativity of
those connectors, neutral elements and reduction rules.

The first invariant we want to maintain is the reduction of singleton for
\emph{seq}, \emph{par\/} and \emph{copar\/} structures. If we try to build a
\verb|cop|, \verb|par| or \verb|seq| with an empty list of structures, then the
creation function shall return the unit \verb|o|. Else if the list contains
only one element, it has to return this element. Otherwise, it will just build
the requested structure. As all manipulated terms are canonical forms, we do not
have for this invariant to handle the case of a structure list containing the
unit, as it will be enforced by the list invariants. This behavior can be
implemented as a \emph{hook\/} for the \verb|seq|, \verb|par| and \verb|cop|
operators.

\begin{verbatim}
par(parl:StrucPar) -> Struc
par:make (l) {
  %match(StrucPar l) {
    concPar() -> { return `o(); }
    concPar(x)-> { return `x; }
  }
  return `make_par(l);
}
\end{verbatim}

This simple \emph{hook\/} implements the invariant for singletons for
\verb|par|, and use a call to the {\tom} constructor \verb|make_par(l)| to call
the intern constructor (without the normalization process), to avoid an
infinite loop. Similar hooks are added to the {\gom} description for \verb|cop|
and \verb|seq| operators. We see here how the pattern matching facilities of
{\tom} embedded in {\gom} can be used to easily implement normalization
strategies.

The \emph{hooks\/} for normalizing structure lists are more complex. They first
require a total order on structures. This can be easily provided as a function,
defined in a \verb|factory| hook. The comparison function we provide here uses
the builtin translation of {\gom} generated data structures to text to
implement a lexical total order. A more specific (and efficient) comparison
function could be written, but for the price of readability. 
\begin{verbatim}
factory {
  public int compareStruc(Object t1, Object t2) {
    String s1 = t1.toString();
    String s2 = t2.toString();
    int res = s1.compareTo(s2);
    return res;
  }
}
\end{verbatim}
Once this function is provided, we can define the hooks for the variadic operators
\verb|concSeq|, \verb|concPar| and \verb|concCop|. The hook for \verb|concSeq|
is the simplest, since the $\seqs{}$ structures are only associative, with $\un$
as neutral element. Then the corresponding hook has to remove the units, and
flatten nested \verb|seq|.
\begin{verbatim}
concSeq( Struc* )  -> StrucSeq
concSeq:make_insert(e,l) {
  %match(Struc e) {
    o()              -> { return l; }
    seq(concSeq(L*)) -> { return `concSeq(L*,l*); }
  }
  return `make_concSeq(e,l);
}
\end{verbatim}
This \emph{hook\/} only checks the form of the element to add to the arguments
of the variadic operator, but does not use the shape of the previous arguments.
The \emph{hooks\/} for \verb|concCop| and \verb|concPar| are similar, but they
do examine also the previous arguments, to perform sorted insertion of the new
argument. This leads to a sorted list of arguments for the operator, providing
a canonical representative for commutative structures.
\begin{verbatim}
concPar( Struc* )  -> StrucPar
concPar:make_insert(e,l) {
  %match(Struc e) {
    o() -> { return l; }
    par(concPar(L*)) -> { return `concPar(L*,l*); }
  }
  %match(StrucPar l) {
    concPar(head,tail*) -> {
      if(!(compareStruc(e, head) < 0)) {
        return `make_concPar(head,concPar(e,tail*));
      }
    }
  }
  return `make_concPar(e,l);
}
\end{verbatim}
The associative matching facility of {\tom} is used to examine the arguments of
the variadic operator, and decide whether to call the builtin construction
function, or perform a recursive call to get a sorted insertion.

As the structure calculus verify the De Morgan rules for the negation, we could
write a hook for the \verb|neg| construction function applying the De Morgan
rules as in Section~\ref{sub:canon} to ensure only atoms are negated. This
will make implementing the deduction rules even simpler, since there is then no
need to propagate negations in the rules.

\subsection{The rules}

Once the data structure is defined, we can implement proof search in system
{\BV} in a {\tom} program using the {\gom} defined data structure by applying
rewriting rules corresponding to the calculus rules to the input structure
repeatedly, until reaching the goal of the prover (usually, the unit {\un}).

Those rules are expressed using {\tom}'s pattern matching over the {\gom} data
structure. They are kept simple because the equivalence relation over
structures is integrated in the data structure with invariants. In this
example, $\pars{}$ and $\aprs{}$ structures are associative and commutative,
while the canonical representatives we use are sorted and flattened variadic
operators.

For instance, the rule $\swir$ of Figure~\ref{fig:BV} can be expressed as the
two rules $\pars{\aprs{R,T},U} \rightarrow \aprs{\pars{R,U},T}$ and
$\pars{\aprs{R,T},U} \rightarrow \aprs{\pars{T,U},R}$, using only associative
matching instead of associative commutative matching.
Then, those rules are encoded by the following match construct, which is placed
into a strategy implementing rewriting in arbitrary context (congruence) to get
deep inference, the \texttt{c} collection being used to gather multiple results:
\begin{verbatim}
%match(Struc t) {
  par(concPar(X1*,cop(concCop(R*,T*)),X2*,U,X3*)) -> {
    if(`T*.isEmpty() || `R*.isEmpty() ) { } 
    else {
      StrucPar context = `concPar(X1*,X2*,X3*);
      if(canReact(`R*,`U)) {
        StrucPar parR = cop2par(`R*);
            // transform a StrucCop into a StrucPar
        Struc elt1 = `par(concPar(
              cop(concCop(par(concPar(parR*,U)),T*)),context*));
        c.add(elt1);       
      }
      if(canReact(`T*,`U)) {
        StrucPar parT = cop2par(`T*);
        Struc elt2 = `par(concPar(
              cop(concCop(par(concPar(parT*,U)),R*)),context*));
        c.add(elt2);  
}  }  } }
\end{verbatim}
We ensure that we do not execute the right-hand side of the rule if 
either \texttt{R} or \texttt{T} are empty lists. The other tests 
implement restrictions on the application of the rules
reducing the non-determinism. 
This is done by using an auxiliary predicate function \verb|canReact(a,b)|
which can be expressed using all the expressive power of both {\tom} and
{\java} in a \verb|factory| hook. The interested reader is referred
to~\cite{KahramanogullariMR-SD05} for a detailed description of those
restrictions.  

Also, the search strategy can be carefully crafted using both {\tom} and
{\java} constructions, to achieve a very fine grained and evolutive strategy,
where usual algebraic languages only allow breadth-first or depth-first
strategies, but do not let the programmer easily define a particular hybrid
search strategy. While the {\tom} approach of search strategies may lead to more
complex implementations for simple examples (as the search space has to be
handled explicitly), it allows us to define fine and efficient strategies for
complex cases.

The implementation of a prover for system {\BV} with {\gom} and {\tom} leads
not only to an efficient implementation, allowing to cleanly separate concerns
about strategy, rules and canonical representatives of terms, but also to
an implementation that can be proven correct, because most parts are expressed
with the high level constructs of {\gom} and {\tom} instead of pure {\java}.
As the data structure invariants in {\gom} and
the deduction rules in {\tom} are defined algebraically, it is possible to
prove that the implemented system is correct and complete with respect to the
original system~\cite{KahramanogullariMR-SD05}, while benefiting from the
expressive power and flexibility of {\java} to express non algebraic concerns
 (like building a web applet for the resulting program, or sending the results
in a network).
%
\section{Conclusion}
\label{sec:conclusion}

We have presented the {\gom} language, a language for describing algebraic
signatures and normalization systems for the terms in those signatures.
This language is kept low level by using {\java} and {\tom} to express the
normalization rules, and by using \emph{hooks\/} for describing how to use the
normalizers. This allows an efficient implementation of the resulting data
structure, preserving properties important to the implementation level, such as
maximal subterm sharing and a strongly typed implementation.
We have shown how this new tool interacts with the {\tom} language. As {\tom}
provides pattern matching, rewrite rules and strategies in imperative languages
like {\C} or {\java}, {\gom} provides algebraic data structures and canonical
representatives to {\java}. Even though {\gom} can be used simply within
{\java}, most benefits are gained when using it with {\tom}, allowing to
integrate formal algebraic developments into mainstream languages. This
integration can allow to formally prove the implemented algorithms with high
level proofs using rewriting techniques, while getting a {\java} implementation
as result.

We have applied this approach to the example of system {\BV} in the structure
calculus, and shown how the method can lead to an efficient implementation for
a complex problem (the implemented prover can tackle more problems than
previous rule based implementation~\cite{KahramanogullariMR-SD05}).

As the compilation process of {\tom}'s pattern matching is formally verified
and shown correct~\cite{KirchnerMR-PPDP2005}, proving the correctness of the
generated data structure and normalizers with respect to the {\gom} description
would allow to expand the trust path from the high level algorithm expressed
with rewrite rules and strategies to the {\java} code generated by the
compilation of {\gom} and {\tom}. This allows to not only prove the correctness
of the implementation, but also to show that the formal parts of the
implementation preserve the properties of the high level rewrite system, such
as confluence or termination.

\noindent
{\bf Acknowledgments:}
I would like to thank Claude Kirchner, Pierre \'Etienne Moreau and all the
{\tom} developers for their help and comments.
Special thanks are due to Pierre Weis and Frederic Blanqui for fruitful
discussions and their help in understanding the design issues.
%
\bibliographystyle{entcs}


\end{document}